\documentclass[aps,prd,twocolumn,showpacs,groupedaddress,nofootinbib]{revtex4}
\usepackage{graphicx}

\begin{document}

\title{Gluonic excitation of non-exotic hybrid charmonium from lattice QCD}

\author{Xiang-Qian Luo\footnote{Email address: stslxq@zsu.edu.cn} and Yan Liu}

\address{Department of Physics, Zhongshan (Sun Yat-Sen) University,
Guangzhou 510275, China}

\date{\today}

\begin{abstract}
The ground and first excited states of the hybrid charmonium
${\bar c} c g$, with non-exotic quantum numbers $J^{PC}=0^{-+}$,
$1^{--}$ and $1^{++}$ are investigated using quenched lattice QCD.
They are completely ignored in the literature, only because their
ground states are degenerate with $\eta_c$, $J/\psi$, and
$\chi_{c1}$, and are difficult to be distinguished from these
conventional charmonium mesons in experiment. However, we observe
strong  gluonic radial excitations in the first excited states; We
predict that their masses are 4.352(225)GeV, 4.379(149)GeV and
7.315(257)GeV, completely different from the first excited states
of the corresponding conventional charmonium. Their relevance to
the recent discovery of the $Y(4260)$ state and future
experimental search for other states are also discussed.
\end{abstract}


\maketitle

A hybrid meson ${\bar q} qg$ is a bound state of constituent quark
$q$, anti-quark ${\bar q}$ and excited gluon $g$. The existence of
hybrids is one of the most important predictions of quantum
chromodynamics (QCD). There has been a lot of experimental
activity\cite{Meyer:2003zq,Peters:2005dg,Olsen:2004yt,Carman:2005ps}
in the search for hybrid mesons, for example: PEP-$\amalg$(Babar),
KEKB(Belle), 12 GeV Jefferson Lab upgraded, upgraded CLEO-c
detector, and new BES3 detector.

For a conventional meson in the quark model, which is represented
by the fermion bilinear ${\bar \psi} \Gamma \psi$, it can have the
$J^{PC}$ quantum numbers as $J=\vert L-S\vert$, $\vert
L-S\vert+1$, ......, $L+S$, $P=(-1)^{L+1}$, and $C=(-1)^{L+S}$,
with $L$ the relative angular momentum of the quark and
anti-quark, and $S$ the intrinsic spin of the meson. For the
gluon, the quantum numbers of the color electric field ${\bf E}$
and color magnetic field ${\bf B}$ are $1^{--}$ and $1^{+-}$
respectively. According to QCD, the operator of a hybrid meson is
the gauge-invariant direct product of the fermion bilinear ${\bar
\psi} \Gamma \psi$ and the color electric field
$E_i^{c_1c_2}=F_{0i}^{c_1c_2}$ or color magnetic field
$B_i^{c_1c_2}=\epsilon_{ijk} F_{jk}^{c_1c_2}$. Therefore, the
quantum numbers of a hybrid meson could be either exotic, with
$J^{PC}=1^{-+}$, $0^{+-}$, $0^{--}$, $2^{+-}$ ......, inaccessible
to conventional mesons, or non-exotic, with $J^{PC}=0^{++}$,
$0^{-+}$, $1^{--}$, $1^{++}$, $1^{+-}$, $2^{++}$, $2^{--}$,
$2^{-+}$, ......, the same as conventional mesons.

Lattice gauge theory is the most reliable technique for computing
hadron spectra. It involves discretization of the continuum theory
on a space-time grid, and reduces to QCD when the lattice spacing
goes to zero. The implementation of the Symanzik
program\cite{Symanzik:1983dc} with tadpole
improvement\cite{Lepage:1992xa} greatly reduces the discretization
errors on very coarse and small lattices. Simulations on
anisotropic lattices improve the signal in spectrum
computations\cite{Mei:2002ip}.

The $1^{-+}$, $0^{+-}$, and $2^{+-}$ exotic hybrid mesons have
been extensively studied on the lattice. Reviews can be found in
Refs.\cite{McNeile:2002en,Michael:2003xg}. Recently, we computed
the $0^{--}$ exotic hybrid charmonium mass\cite{Liu:2005rc}.
However, the non-exotic hybrid mesons are usually ignored in the
literature, simply because their ground states are almost
degenerate with the conventional mesons\cite{Liao:2002rj}.

In this letter, we investigate the $J^{PC}=0^{-+}$, $1^{--}$ and
$1^{++}$ non-exotic charmed hybrid mesons ${\bar c} cg$, employing
quenched lattice QCD with tadpole improved
gluon\cite{Morningstar:1997prd} and quark\cite{Okamoto:2001jb}
actions on the anisotropic lattice. We observe, for the first
time, very strong gluonic radial excitations in the first excited
states.

\begin{table*}
\begin{center}
\tabcolsep 0.1in
\begin{tabular}{cccccccccccc}\hline
 $\beta$ & $\xi=a_s/a_t$ & $L^{3}\times T$   & $u_{s}$  & $u_t$  & $a_{t}m_{q0}$                   & $c_{s}$ & $c_{t}$ & $a_{s}$($1^{1}P_{1}-1S$)[fm]  & $La_{s}$[fm] & \\\hline
    2.6  &    3          & $16^{3}\times 48$ & 0.81921  &  1     & 0.229 $\tabcolsep 0.2in$ 0.260  & 1.8189  & 2.4414  & 0.1856(84)                    &  2.970       & \\
    2.8  &    3          & $16^{3}\times 48$ & 0.83099  &  1     & 0.150 $\tabcolsep 0.2in$ 0.220  & 1.7427  & 2.4068  & 0.1537(101)                   &  2.459       & \\
    3.0  &    3          & $20^{3}\times 60$ & 0.84098  &  1     & 0.020 $\tabcolsep 0.2in$ 0.100  & 1.6813  & 2.3782  & 0.1128(110)                   &  2.256       & \\
\hline
\end{tabular}
\end{center}
\caption{Simulation parameters at largest volume. We employed the
method in Ref.\cite{Okamoto:2001jb} to tune these parameters,
$\kappa_{t}$ and $\kappa_{s}$ for the quark action. The last two
columns are about the spatial lattice spacing and the lattice
extent in physical units, determined from the $1P-1S$ charmonium
mass splitting\cite{Liu:2005rc}, with the effective masses
extracted by the method of Ref. \cite{Bernard:2003jd}.}
\label{tab1}
\end{table*}

Our simulation parameters are listed in Tab. \ref{tab1}. We also
did simulations on $8^3\times48$ and $12^3\times 48$ at
$\beta=2.6$, $12^3\times36$ at $\beta=2.8$, and $16^3\times 48$ at
$\beta=3.0$, but there and throughout the paper we just list the
results from the largest volume, i.e., $16^{3}\times 48$ at
$\beta=2.6$ and $\beta=2.8$ and $20^3\times 60$ at $\beta=3.0$. At
each $\beta=6/g^2$, three hundred independent configurations were
generated with the improved gluonic
action\cite{Morningstar:1997prd}. It is also important to check
whether these lattice volumes are large enough. When the spatial
extent is greater than 2.2fm, the finite volume effect on the
spectrum is less than $0.1\%$ for the ground state, and $0.4\%$
for the first excited state.

\begin{figure} [th]
\begin{center}
\includegraphics[totalheight=2.5in]{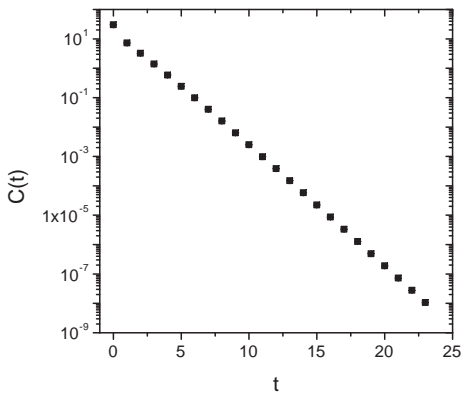}
\end{center}
\vspace{0.0cm} \caption{Correlation function for the conventional
$1^{--}$ quarkonium at $\beta=2.6$ and $a_{t}m_{q0}=0.229$.}
\label{fig1}
\end{figure}

\begin{figure} [th]
\begin{center}
\includegraphics[totalheight=2.5in]{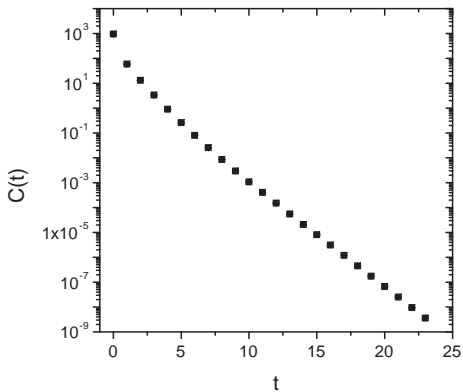}
\end{center}
\vspace{0.0cm} \caption{Same as Fig. \ref{fig1}, but for the
$1^{--}$ hybrid meson.} \label{fig2}
\end{figure}

We input the bare quark mass $m_{q0}$ and then computed quark
propagators using the improved quark action\cite{Okamoto:2001jb},
the conventional quarkonium correlation function using the
operators $0^{-+}=\bar{\psi}^{c}\gamma_{5}\psi^{c}$,
$1^{--}=\bar{\psi}^{c}\gamma_{j}\psi^{c}$, and
$1^{++}=\bar{\psi}^{c}\gamma_5\gamma_{j}\psi^{c}$, and the hybrid
meson correlation function using the operators
$0^{-+}=\epsilon_{ijk}\bar{\psi}^{c_1}\gamma_{i}\psi^{c_2}F^{c_1c_2}_{jk}$,
$1^{--}=\epsilon_{ijk}\bar{\psi}^{c_1}\gamma_{5}\psi^{c_2}F^{c_1c_2}_{jk}$
and
$1^{++}=\epsilon_{ijk}\bar{\psi}^{c_1}\gamma_{j}\psi^{c_2}F^{c_1c_2}_{0k}$
in Ref. \cite{Bernard:1997ib}. Figures \ref{fig1} and \ref{fig2}
shows the correlation function $C(t)$ of the conventional $1^{--}$
and hybrid mesons.

The effective masses of the ground and first excited states
$a_tm_1$ and $a_tm_2$ are extracted by two different methods: (i)
new correlation function method\cite{Guadagnoli:2004wm}; (ii)
modified multi-exponential fit\cite{Bernard:2003jd}. The
multi-exponential fitting method has been widely used in the
literature\cite{Okamoto:2001jb,Liao:2002rj,Chen:2000ej} for
extracting the charmonium masses, and the results for the ground
and first excited states are consistent with experiments; The MILC
group\cite{Bernard:2003jd} proposed an improved multi-exponential
fitting method, which chooses the best fit according to some
criteria. The recently proposed method (i) has been successfully
applied to the investigation of the Roper resonance of the
nucleon\cite{Guadagnoli:2004wm}, where $a_t m_1$ is obtained from
$\ln(C(t)/C(t+1))$ in the large time interval $[t_i,t_f]$, and
$a_t m_1+ a_tm_2$ from $\ln(C'(t)/C'(t+1))$ in the time interval
$[t_i^*,t_f^*]<[t_i,t_f]$, with reasonable $\chi^2/d.o.f.$ and
optimal confidence level. Here $C'(t)=C(t+1)C(t-1)-C(t)^2$. Two
methods provide a cross-check of the results.

\begin{figure} [th]
\begin{center}
\includegraphics[totalheight=2.5in]{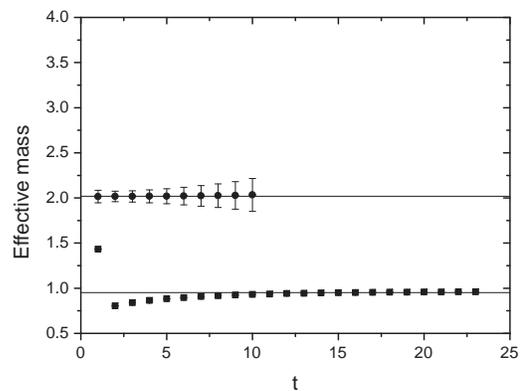}
\end{center}
\vspace{0.0cm} \caption{Effective masses of the conventional
$1^{--}$ quarkonium as a function of $t$ for $\beta=2.6$ and
$a_{t}m_{q0}=0.229$, using the new correlation function
method\cite{Guadagnoli:2004wm}. $a_t m_1 + a_t m_2$ and $a_t m_1$
are extracted respectively from the plateaux of the upper and
lower curves, with $[t_i^*,t_f^*]=[1,10]$ and
$[t_i,t_f]=[11,23]$.} \label{fig3}
\end{figure}

\begin{figure} [th]
\begin{center}
\includegraphics[totalheight=2.5in]{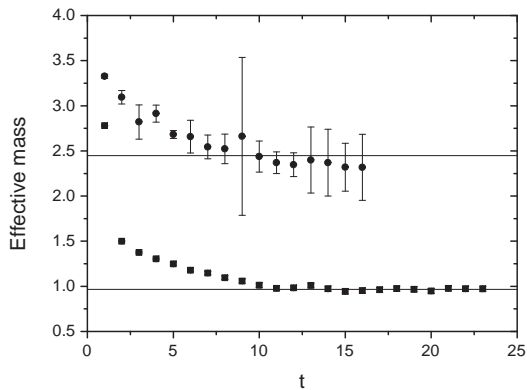}
\end{center}
\vspace{0.0cm} \caption{Same as Fig. \ref{fig3}, but for the
$1^{--}$ hybrid meson. $a_t m_1 + a_t m_2$ and $a_t m_1$ are
extracted respectively from the plateaux of the upper and lower
curves, with $[t_i^*,t_f^*]=[6,16]$ and $[t_i,t_f]=[17,23]$.}
\label{fig4}
\end{figure}

Figure \ref{fig3} shows effective masses for the conventional
$1^{--}$ quarkonium, where $a_t m_1$ and $a_t m_1 + a_t m_2$ are
extracted respectively from the plateaux of the lower and upper
curves, using the new method\cite{Guadagnoli:2004wm}. Figure
\ref{fig4} shows those for the $1^{--}$ hybrid meson.

\begin{figure} [th]
\begin{center}
\includegraphics[totalheight=2.5in]{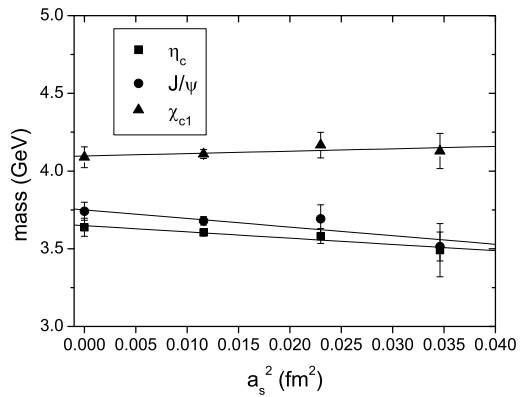}
\end{center}
\caption{Extrapolation of the excited state masses of  the
conventional $0^{-+}$, $1^{--}$, and $1^{++}$ charmonium mesons,
with the effective masses extracted by the method of Ref.
\cite{Guadagnoli:2004wm}, to the continuum limit.} \label{fig5}
\end{figure}

\begin{figure} [th]
\begin{center}
\includegraphics[totalheight=2.5in]{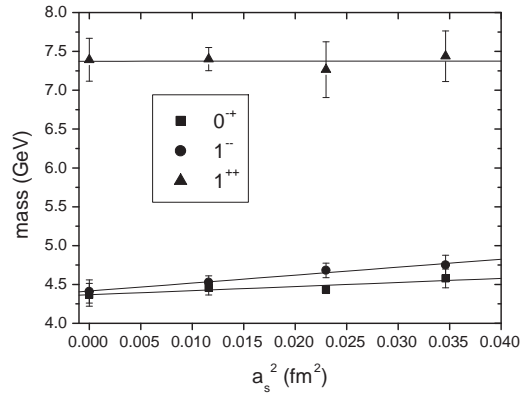}
\end{center}
\caption{The same as Fig. \ref{fig5}, but for the hybrid
charmonium mesons.} \label{fig6}
\end{figure}

\begin{table*}
\begin{center}
\tabcolsep 0.15in
\begin{tabular}{ccccccccc}\hline
 $\beta$   & $a_{s}^{2}(fm^{2})$ &$\eta_c$     & $J/\psi$   &$\chi_{c1}$   & $0^{-+}$  & $1^{--}$  & $1^{++}$ &~\\\hline
    2.6    & 0.0345              &3.031(3)     &3.080(3)    &3.484(49)      &3.012(43)  &3.133(44)  &3.472(66)&~\\
    2.8    & 0.0236              &3.033(1)     &3.079(1)    &3.446(59)      &3.009(51)  &3.112(53)  &3.463(62)&~\\
    3.0    & 0.0127              &3.031(1)     &3.080(1)    &3.458(64)      &3.027(57)  &3.099(58)  &3.506(65)&~\\
$\infty$   & 0                   &3.030(2)     &3.080(2)    &3.430(100)     &3.031(87)  &3.078(90)  &3.516(108)&~\\
\hline
$\infty$   & 0                   &3.053(33)    &3.107(34)   &3.533(39)      &3.056(34)  &3.120(34)  &3.472(150)&* \\
\hline \hline
$\infty$   & 0                   &3.042(18)    &3.094(18)   &3.482(70)      &3.044(61)  &3.099(62)  &3.494(129)&** \\
 \hline
\end{tabular}
\end{center}
\caption{Conventional and hybrid charmonium meson spectrum (GeV)
for the ground state from the method of Ref.
\cite{Bernard:2003jd}, interpolated to the charm quark sector. The
results in the continuum limit ($\beta=\infty$) were obtained by
linearly extrapolating the data to $a_s^2 \to 0$. The results in
the continuum limit(*), with the effective masses extracted by the
method of Ref. \cite{Guadagnoli:2004wm} are also listed. The last
line (**) is the average of the results in the continuum limit
from these two methods.} \label{tab2}
\end{table*}

\begin{table*}
\begin{center}
\tabcolsep 0.15in
\begin{tabular}{ccccccccc}\hline
$\beta$ & $a_{s}^{2}(fm^{2})$  &$\eta_c$ & $J/\psi$ & $\chi_{c1}$
& $0^{-+}$  & $1^{--}$   & $1^{++}$ & ~ \\\hline
    2.6    & 0.0345               &3.515(50)    &3.614(51)   &4.135(175)    &4.492(64)   &4.525(64)   &7.335(121) &~   \\
    2.8    & 0.0236               &3.520(60)    &3.625(62)   &4.175(183)    &4.379(98)   &4.494(77)   &7.333(153) &~   \\
    3.0    & 0.0127               &3.532(66)    &3.624(68)   &4.100(112)    &4.408(382)  &4.400(100)  &7.264(150) &~   \\
$\infty$   & 0                    &3.540(102)   &3.633(105)  &4.081(205)    &4.335(302)  &4.349(148)  &7.237(237) &~   \\
\hline
$\infty$   & 0                  &3.638(58)     &3.731(58)   &4.089(67)     &4.368(147)  &4.409(149)  &7.392(276) &*   \\
\hline \hline
$\infty$   & 0                  &3.589(80)     &3.682(81)   &4.085(136)     &4.352(225)  &4.379(149)  &7.315(257) &**   \\
\hline
\end{tabular}
\end{center}
\caption{The same as Tab. \ref{tab2}, but for the first excited
states.} \label{tab3}
\end{table*}

The data at two $m_{q0}$ values were interpolated to the charm
quark regime using
$M(1S)_{exp}=(m(\eta_{c})_{exp}+3m(J/\psi)_{exp})/4 = 3067.6$MeV.
The results obtained by the method of Ref. \cite{Bernard:2003jd}
are listed in Tabs. \ref{tab2} and \ref{tab3} respectively for the
ground and first excited states.

The first excited state masses for the conventional $0^{-+}$,
$1^{--}$ and $1^{++}$ charmonium mesons as a function of $a^2_s$
are plotted in Fig. \ref{fig5}, and those for the hybrid
charmonium mesons are plotted in Fig. \ref{fig6}. They indicate
the linear dependence of the mass on $a^2_s$. By linearly
extrapolating the data to $a_s^2\to 0$, we obtained the spectrum
in the continuum limit, which are listed in Tabs. \ref{tab2} and
\ref{tab3}.

As shown in Tab. \ref{tab2}, in the continuum limit, the masses of
the $0^{-+}$, $1^{--}$ and $1^{++}$ charmonium ground states are
consistent with their experimental values 2.9804, 3.0969, and
3.5106 for $\eta_c(1S)$, $J/\psi$ and $\chi_{c1}(1^3P_1)$. The
results in Tab. \ref{tab2} also show that the ground state for the
non-exotic hybrid charmonium is degenerate with the conventional
charmonium with the same quantum numbers. This might mislead
people into giving up further study of the non-exotic hybrids.

The last line of Tab. \ref{tab3} shows in the continuum limit the
first excited state masses of the conventional charmonium and
non-exotic hybrid charmonium. The results for the conventional
$0^{-+}$ and $1^{--}$ charmonium are in good agreement with the
experimental data 3.638 and 3.686 for $\eta_c(2S)$ and $\psi(2S)$,
which supports the reliability of the methods. Although there has
not been experimental input for $\chi_{c1}(2^3P_1)$, our result is
consistent with earlier lattice
calculations\cite{Okamoto:2001jb,Chen:2000ej}.

The minor differences between the data and experiments might be
due to the quenched approximation used in the paper.

What new is that the first excited states of non-exotic charmonium
hybrids are completely different from the conventional ones. The
results in last line of Tab.\ref{tab3} show the masses of the
$0^{-+}$ and $1^{--}$ hybrids to be  about 0.7GeV heavier, and the
$1^{++}$ about 3.2GeV heavier. These are very strong indications
of gluonic excitations. This implies that radial excitations of
the charmonium hybrids are completely different from the
conventional non-hybrid ones, although their ground states
overlap. This is clearly demonstrated in Figs.
\ref{fig1}-\ref{fig6}.

This also teaches a very important lesson. One should carefully
study not only the ground state, but also the excited states.
Sometimes, the excited states show more fundamental properties of
a hadron.

Finally, we discuss the new state $Y(4260)$, recently observed by
the BaBar experiment\cite{Aubert:2005rm} in the $J/\psi \pi^+
\pi^-$ channel. It has the quantum numbers $J^{PC}=1^{--}$.

There have been several phenomenological
descriptions\cite{Zhu:2005hp,Llanes-Estrada:2005hz,Kou:2005gt,Maiani:2005pe,Liu:2005ay,Close:2005iz,Qiao:2005av}
of this state: as tetra-quarks, a molecule of two mesons,
$\psi(4S)$, or as a hybrid meson; However, most these assumptions
were not based on QCD spectrum computations.

If $Y(4260)$ is a hybrid meson, from the last line of Tab.
\ref{tab2}, it could certainly not be identified as the ground
state of the $1^{--}$ hybrid meson. However, from our lattice QCD
spectrum calculations (the last line of Tab. \ref{tab3}), it is
most probably the first excited state of the $1^{--}$ hybrid
charmonium. Further experimental study of the decay modes will
clarify this issue.

From the last line of Tab. \ref{tab3}, one sees that the first
excited state mass of the $0^{-+}$ hybrid charmonium is about the
same as that of the $1^{--}$ hybrid charmonium, but much lighter
than the first excited state of the $1^{++}$ hybrid charmonium. It
should not be very difficult to find it in future experiment.

\acknowledgments

We thank K.T. Chao, C. DeTar, E.B. Gregory, F.J. Llanes-Estrada,
E. Swanson, D. Toussaint, C.Z. Yuan and S.L. Zhu for useful
discussions. This work is supported by the Key Project of National
Science Foundation (10235040), Project of the Chinese Academy of
Sciences (KJCX2-SW-N10) and Key Project of National Ministry of
Eduction (105135) and Guangdong Natural Science Foundation
(05101821). We modified the MILC code\cite{Milc} for simulations
on the anisotropic lattice. The simulations had taken in total one
year and a half on our AMD-Opteron cluster and Beijing LSSC2 XEON
cluster.

\end{document}